# Theoretical realization of Mo$_2$P; a novel stable 2D material with superionic conductivity and attractive optical properties

Bohayra Mortazavi[*,1], Masoud Shahrokhi[2], Meysam Makaremi[3], Timon Rabczuk[#,4]

[1]*Institute of Structural Mechanics, Bauhaus-Universität Weimar, Marienstr. 15, D-99423 Weimar, Germany.*

[2]*Institute of Chemical Research of Catalonia, ICIQ, The Barcelona Institute of Science and Technology, Av. Països Catalans 16, ES-43007 Tarragona, Spain.*

[3]*Chemical Department of Materials Science and Engineering, University of Toronto, 184 College Street, Toronto, Canada.*

[4]*College of Civil Engineering, Department of Geotechnical Engineering, Tongji University, Shanghai, China.*

## Abstract

Mo$_2$P as a new member of the advancing two-dimensional (2D) materials family has been theoretically identified in this study. We conducted extensive density functional theory calculations to explore the crystal structure, dynamical stability, mechanical response, electronic structure and optical properties. Mo$_2$P was found to be metallic with the Fermi energy locating at the *d* bands of transition metal Mo. A high reflectivity of ~100% at low energies less than 1 eV was observed, introducing Mo$_2$P as a potential candidate for photonic and optoelectronic applications such as transmitting electromagnetic waves devices. Our calculations confirm that the novel 2D structure is dynamically stable and can withstand at high temperatures including 1000 K. Mo$_2$P was found to yield high tensile strength and elastic modulus of 12 GPa.nm and 56 GPa.nm, respectively. We particularly evaluated the application of Mo$_2$P as an anode material for Li and Na-ion rechargeable batteries. The open-circuit voltages of 0.88–1.06 V and 0.94–0.09 V were predicted for Li and Na ions storages, respectively, which are desirable for commercial anodic materials. Interestingly, our calculations predict remarkably low diffusion energy barriers of 50 meV and 17 meV for Li and Na adatoms, respectively, promising to achieve ultrafast charging/discharging. The findings provided by this study can motivate further experimental and theoretical studies to probe new 2D crystals made from phosphor and transition metals with 2H and 1T atomic structures.

*Corresponding author (Bohayra Mortazavi):  bohayra.mortazavi@gmail.com
Tel: +49 157 8037 8770; Fax: +49 364 358 4511

# 1. Introduction

Recently, two-dimensional (2D) structures, a new class of materials has been attracted remarkable attention for numerous applications ranging from nanoelectronics to aerospace devices. The successful mechanical exfoliation of graphene from graphite in 2004 [1,2], raised an ongoing interest toward the synthesis of other fascinating 2D materials with the high quality and large surface area. During the last decade a wide variety of 2D materials including hexagonal boron-nitride [3,4], graphitic carbon nitride [5,6], silicene [7,8], germanene [9], transition metal dichalcogenides [10–14], transition metal oxides [15], phosphorene [16,17] and most recently borophene [18] with outstanding properties have been synthesized. An interesting fact about the 2D materials is that their properties can be finely tuned by chemical functionalization [19–25], defect engineering [26,27] or mechanical loading [28–30]. In the other hand, prior to the experimental synthesis of novel 2D materials, their existence, stability and physical/chemical properties could be analyzed by the state of the art theoretical framework including the density functional theory (DFT) calculations.

In recent years, developments of rechargeable metallic ion batteries technology plays as crucial role for the progress in electronic, communication and automobile industries. In this regard, for particular application like electric vehicles, there exist continuous demands for the faster charging/discharging rates and higher charge capacities. Currently, graphite is the mostly employed anode materials in Li-ion batteries. However, the ion diffusion rate and the charge capacity of graphite are limited. Therefore the need for the substitution of graphite with more advanced materials is inevitable in order to improve the performance efficiency of rechargeable batteries. In the other hand, 2D materials and their hybrid structures owing to their large surface area and remarkable thermal and mechanical stabilities are currently considered as promising solutions to reach high efficient rechargeable batteries. Previous theoretical studies have confirmed that 2D materials and their heterostructures can yield remarkably high capacities [31–33] and low diffusion energy barriers [34–39].

Recently, bulk molybdenum phosphide has attracted attention due to presenting topological semimetal properties and its three-component unconventional fermions [40]. Motivated by this recent experimental advances, we conducted DFT calculations to examine the stability, mechanical, optical and anodic properties of 2D $Mo_2P$. Our first-principles results propose the $Mo_2P$ as a thermodynamically stable and electrically conductive anode materials to achieve ultrafast charging/discharging. Moreover, remarkably high reflectivity of Mo$_2$P at energies lower than 1 eV is promising for the application in photonic and optoelectronic applications. The insight provided by this study can be useful for future experimental and theoretical studies concerning the new 2D materials made from phosphor and transition metals with various atomic structures.

## 2. Computational methods

The structural properties, adatom adsorptions, mechanical response, electronic band structure and optical calculations are computed using the DFT method implemented in the Vienna ab initio simulation package [41–43]. The DFT calculations are within the framework of generalized gradient approximation (GGA) in the form of Perdew-Burke-Ernzerhof (PBE) [44] for the exchange correlation potential and the ion-electron interaction is treated using the projector augmented wave (PAW) method. Electronic wave functions have been expanded using a plane-wave basis set with a cut-off energy of 500 eV. To evaluate the electronic properties, we also used the screened hybrid functional, HSE06 [45]. The interaction of Mo$_2$P with Li or Na adatoms was studied by conducting spin polarized DFT simulations. The van der Waals DFT-D2 correction method of Grimme [46] was also applied to modify the binding energy calculations. We applied periodic boundary conditions along all three Cartesian directions and in order to avoid image-image interactions, the simulation box size along the sheet normal direction was considered to be 20 Å. The electron self-consistent convergence criterion of $1\times10^{-4}$ eV and $1\times10^{-5}$ eV were considered for adatoms adsorption and structural relaxation, respectively. For the ionic relaxation, the conjugate gradient method was employed with the force convergence criterion of $1\times10^{-2}$ eV/Å and $3\times10^{-3}$ eV/Å for adatoms adsorption and structural relaxation, respectively. To simulate the mechanical properties we used a 15×15×1 Monkhorst-Pack [47] k-point mesh size. To simulate the adatoms adsorption, a 4×4 supercell was used in which a 4×4×1 Monkhorst-Pack [5] k-point sampling was employed. In this case after the achievement of energy minimization, a single point calculation was performed to report the energy values as well as the electronic density of states (DOS) using a finer Monkhorst-Pack mesh size of 9×9×1. A 14×14×1 Γ centered Monkhorst-Pack k-point mesh was also used to report the electronic properties within PBE and HSE06.

The thermal stability of single-layer Mo$_2$P was explored via using the ab-initio molecular dynamics (AIMD) simulations for a 5×5 supercell. For the AIMD simulation, the NVT ensemble with a 1 fs time increment and 2×2×1 k-point mesh size was used. To assess the dynamical stability, phonon dispersion was calculated for a 4×4 supercell calculations using the finite displacement method, as implemented in the Phonopy package [48]. The diffusion of a single Li or Na adatom over the single-layer Mo$_2$P and the corresponding energy barrier were calculated using the climbing nudged elastic band (NEB) [49] method. For NEB calculations, we used a 3×2 supercell of the Mo$_2$P.

Optical calculations were performed through the random phase approximation (RPA) method constructed over PBE using full potential Wien2k code [50]. Since we found that strained Mo$_2$P systems present metallic electronic character, the intraband transition contribution was also included to the interband transitions [51] to accurately study the optical properties. The Wien2k package provides the possibility to add the intraband transition contribution to the interband ones. In order to achieve energy eigenvalues convergence in optical calculations, the wave functionals in the interstitial region were expanded in terms of plane waves with a cut-off parameter of $R_{MT} \times K_{max} = 8$, where $R_{MT}$ denotes the smallest atomic sphere radius and $K_{max}$ indicates the largest $k$ vector in the plane-wave expansion. The maximum angular momentum of the atomic orbital basis functions was set to $l_{max}=12$ and the Fourier expansion charge density is truncated at $G_{max} = 14$ Ry$^{1/2}$. For the simulations involving optical properties, Brillouin zone was sampled by 24×24×1 grids and setting Lorentzian broadening with gamma equal to 0.05 eV (more details of optical properties calculations are included in supporting information).

## 3. Results and discussions

Single-layer and free-standing Mo$_2$P present a hexagonal lattice with symmetry group of P6m2 which is composed of a layer of P atoms between two Mo layers. In this study, we explored the properties of 2H Mo$_2$P, which shows the atomic stacking sequence of ABA. Likely to MoS$_2$ [52,53], the Mo$_2$P may also show 1T phase with ABC stacking sequence, in which the Mo atoms on the bottom are placed in the hollow center of the hexagonal lattice. Fig.1, illustrates the atomic structure of single-layer Mo$_2$P, which shows the hexagonal lattice constant and Mo-P bond length of 3.064 Å and 2.347 Å, respectively. The electron localization function (ELF) [54] is also illustrated in Fig. 1c. As it is clear, electron localization occur between the Mo-P bonds, which confirms the covalent bonding. Nevertheless, the charge localization is more concentrated around the P atoms as compared with the Mo atoms. For the more discerning of the bonding characters between Mo and P atoms, we conducted the Bader charge analysis [55]. It was found that on the average, ~0.35 electrons transfer from each Mo to P atom. In another word, each Mo atom losses ~0.35 electrons and each P atom gains ~0.70 electrons. After combining the ELF and Bader analysis results, it can be concluded that 2D $Mo_2P$ can be classified as an ionic-covalent crystal with a small charge transferring from Mo to P atoms.

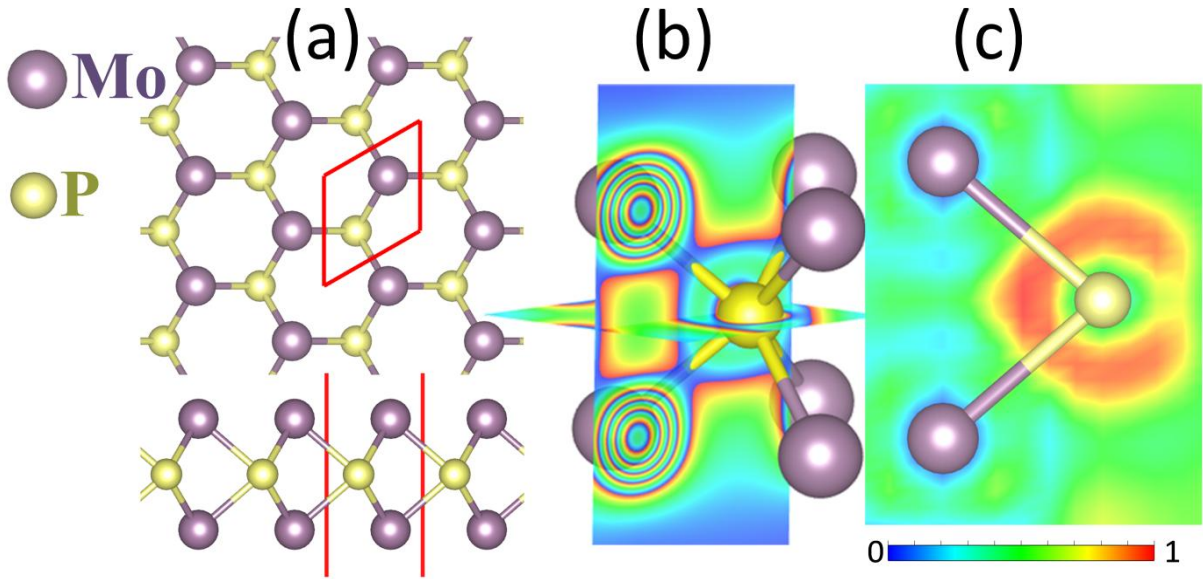

Fig. 1, (a) Top and side views of the atomic configuration, (b) electron charge density and (c) electron localization function (ELF) of the single-layer $Mo_2P$ (see SI for detailed information of $Mo_2P$ lattice). ELF has a value between 0 and 1, where 1 corresponds to perfect localization.

To examine the energetic stability of $Mo_2P$, the cohesive energy per atom was calculated as defined by; $E_{coh} = (\sum_i E_i - E_t)/n$, where $E_t$, $E_i$ and $n$ denote the total energy per cell, the energy of the *i-th* isolated atom and the total number of atoms per cell, respectively. $E_{coh}$ of energy minimized single-layer $Mo_2P$ was obtained to be 9.95 eV, which confirms the energetic stability of the structure. To further explore the dynamical stability of the $Mo_2P$ single-layer, the phonon dispersion along the high symmetry Γ-M-K-Γ directions were calculated, and depicted in Fig. 2a in which there exist no imaginary vibrational frequency in the phonon dispersion; confirming the stability of the structure. Moreover, AIMD simulation were performed at different temperatures for a simulation time of 10 ps. As can be seen in the AIMD snapshot for a sample at 1000 K (Fig. 2b), the structure is stable and there exist no persistent damage or bond rupture in the lattice. The obtained results confirm the remarkable energetic, dynamical and thermal stability of 2D $Mo_2P$ and such that synthesize of this 2D material in real experimental conditions should be achievable.

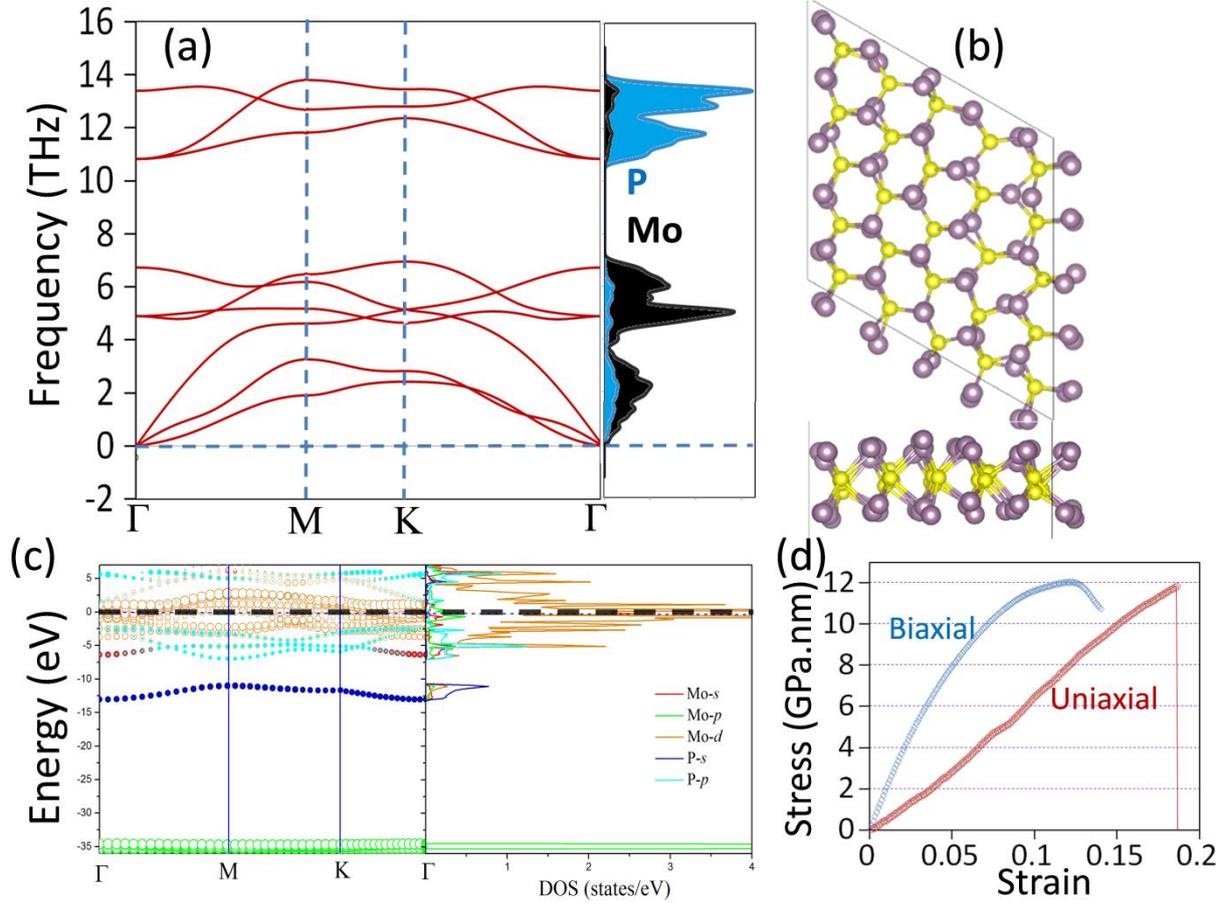

Fig-2, (a) Phonon dispersions of free-standing $Mo_2P$ does not show negative frequencies which confirms its dynamical stability. (b) Top and side views of a snapshot of 2H $Mo_2P$ at 1000 K, reveal the remarkable thermal stability of this system. (c) Partial band structure and DOS of $Mo_2P$ monolayer obtained using the PBE approach illustarting the metallic electronic nature of $Mo_2P$. (d) Stress-strain response of single-layer $Mo_2P$ under biaxial or uniaxial loading conditions.

To probe the electronic properties of free-standing $Mo_2P$, the partial electronic band structure along the high symmetry Γ-M-K-Γ directions and partial density of states (DOS) were calculated. Figs. 2c illustrates the total band structure and DOS of the $Mo_2P$ monolayer based on the PBE approach. The calculations show that stress-free

Mo$_2$P sheet is metallic with the Fermi energy locating at the $d$ bands of transition metal Mo. The $p$ bands of P atom appear below the $d$ bands of Mo atom and the $s$ bands of the P atom are located below the $p$ bands of the P atom separated by a band gap. A strong hybridization of Mo 4$d$ and P 3$p$ states around the Fermi energy is also conspicuous.

Next, we analyze the mechanical and electronic responses of single-layer Mo$_2$P by performing uniaxial and biaxial loading conditions. For the biaxial loading, the structure was strained equally along both of the planar directions. For the sample under the uniaxial stress condition, the loading strain was only applied along the loading direction. In this case, the simulation box size along the perpendicular direction of the loading was altered in a way that after the energy minimization the stress on the perpendicular direction of the loading remains negligible and accordingly satisfy the uniaxial stress condition [56,57]. Based on the stress-strain results shown in Fig. 2d, the Mo$_2$P can yield a high tensile strength of around 12 GPa.nm under the biaxial or uniaxial loading. We note that the initial linear region of the uniaxial stress-strain is equivalent with the elastic modulus. This way, the elastic modulus of single-layer Mo$_2$P was calculated to be 56 GPa.nm. Our results for the biaxial and uniaxial loading clearly confirm remarkable mechanical stiffness of Mo$_2$P. In this study, we explored the evolution of optical and electronic properties for the biaxial and uniaxial loadings at five different strain levels with respect to the strain at the tensile strength point, $s_{uts}$. Table S1, summarizes the cohesive energy of Mo$_2$P monolayer under different magnitudes of strain. The results exhibit that the cohesive energy for all strained systems is positive confirming the energetic stability of these structures. Nonetheless, by increasing the magnitude of the strain, the cohesive energy decreases for both of the uniaxial and biaxial loadings. We also analyzed the total band structure and DOS for the Mo$_2$P nanomembranes at different magnitudes of strain by using the PBE approach (as shown in Fig. S2). According to the band structure and DOS results for the stretched samples, the conduction and valence bands overlap at Fermi level demonstrating a metallic characteristic. We have also used HSE06 approaches to compute the total DOS for stress-free and strained Mo$_2$P (Fig. S3). In accordance with the PBE results, the HSE06 approach doesn't reveal any band gap opening in single-layer Mo$_2$P upon the biaxial or uniaxial stretching.

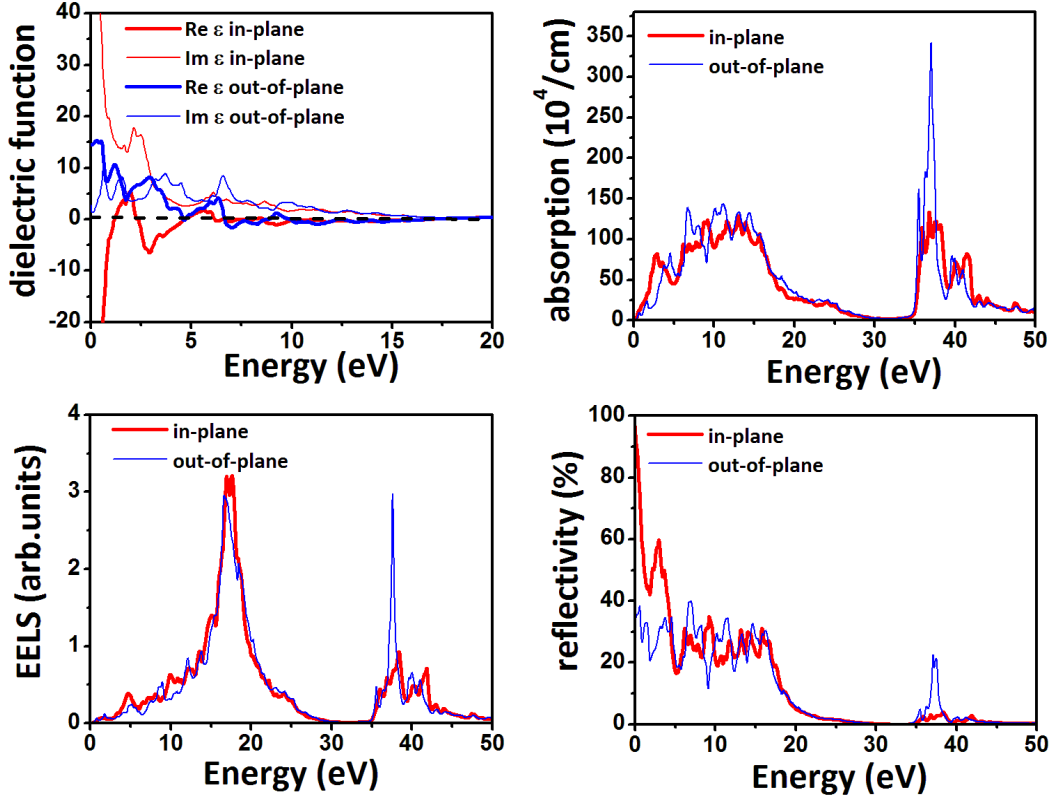

Fig. 3, Calculated in-plane and out-of-plane optical spectra of $Mo_2P$ monolayer by considering the intraband transitions contribution.

We next shift our attention to the optical properties of single-layer $Mo_2P$. The optical properties of these materials have been studied using the RPA [58] method by taking into account the contribution of inter and intraband transitions. It is found that the optical spectra become anisotropic along in-plane (E||x or y) and out-of-plane (E||z) directions. By considering the intraband transition contribution, the imaginary and real parts of the dielectric function have a singularity at zero frequency along in-plane direction, presents optically metallic property. In contrast, the positive values of static dielectric constant (the values of Re ε at zero energy) and nearly zero first plasma frequency confirm the semiconductor characters for optical spectra along out-of-plane direction (for more information see SI). The reason of this behavior is the huge depolarization effect along out-of-plane direction for 2D materials. These properties have also been seen for other metallic and semi-metallic 2D sheets like $B_2C$ [59], $Mo_2C$ [56], MXene monolayers [60] and graphene [61,62].

Fig. 3 depicts the dielectric function, absorption coefficient, electron energy loss spectra (EELS) and reflectivity for Mo2P monolayer. The first plasma frequency of $Mo_2P$ monolayer occurs at energy 4 eV when the electric field is applied parallel to the surface which is related to π electron plasmon peak. The main plasma peak is broad and occurs around 17 eV for in-plane direction that is related to π + σ electron plasmon peak. The first and main plasma frequencies for in-plane direction shift to lower energies by applying strain to the system (Table S1). The first absorption peak for in-plane direction occurs at visible range (~3 eV), which renders its potential applications in optical and electronic devices. At energy range between 25 and 35 eV, reflectivity and absorption of $Mo_2P$ sheet is nearly zero for in-plane and out-of plane direction where the transmission is maximum. Furthermore, at extremely low energy less than 1 eV, the reflectivity becomes ~100% (less than 40%) for these systems in parallel (perpendicular) polarization. This indicates the capability of these systems for transmitting electromagnetic waves. The similar reflectivity properties have also been seen for MXene monolayers [63]. The effects of different loading conditions and strains on the optical properties have been analyzed elaborately in the supporting information.

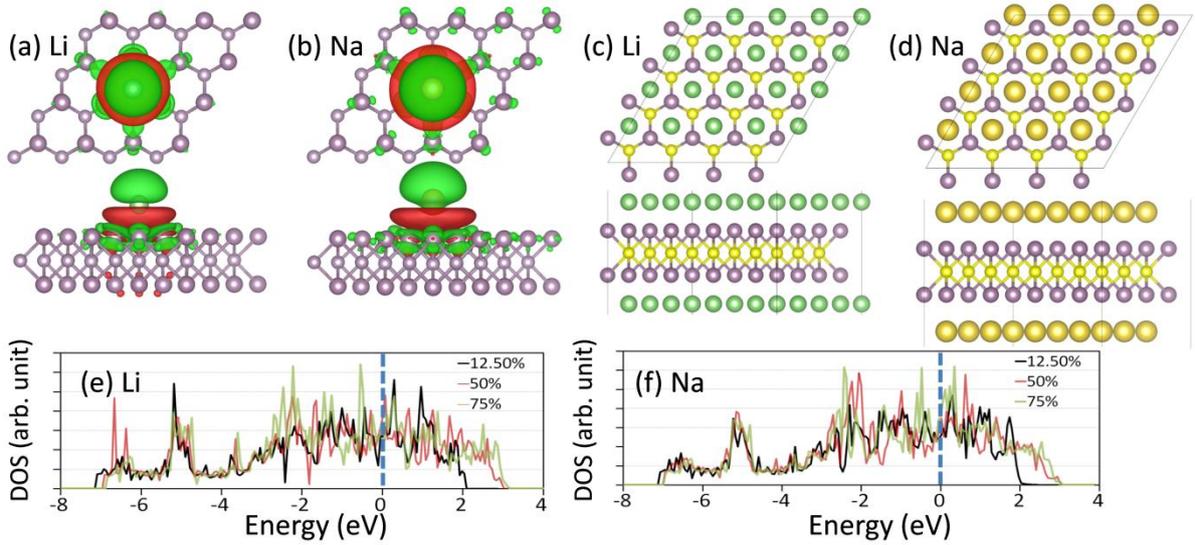

Fig. 4, The differential charge density of the single (a) Li or (b) Na adatom over $Mo_2P$ monolayer. Color coding consists of red for charge gain and green for charge loss. Top and side views of monolayer $Mo_2P$ with maximum coverage of (c) Li and (d) Na adatoms. Total density of states for the $Mo_2P$ monolayer covered with various concentration of (e) Li and (f) Na adatoms ensure the good electronic conductivity.

At the final step, we investigated the application of $Mo_2P$ for the Li and Na storage. The adsorption energy of metallic adatoms is a key factor in determining a material as a proper anodic candidate. In this case, we first identified the strongest adsorption site for each adatom binding on the $Mo_2P$ monolayer. The adsorption energy, $E_{ad}$, can be calculated by:

$$E_{ad} = E_{AM} - E_A - E_M \qquad (1)$$

where $E_A$ and $E_{AM}$ are respectively, the total energies of $Mo_2P$ before and after the metal adatoms adsorption and $E_M$ is the per atom lattice energy of the metallic adatom. Three different adsorption sites were considered; over hexagonal hollow, top of the P atoms and top of the Mo atoms. For both Li and Na adatoms, we found that hexagonal hollow sites present the maximum binding energy. The adsorption energy of single Li and Na adatoms interacting with the $Mo_2P$ single-layer were calculated to be -0.87 eV and -1.00 eV, respectively. In Figs. 4a and 4b, top and side views of single Li and Na atom adsorptions on the hexagonal hollow site of monolayer $Mo_2P$ are shown. Here, we also plotted the charge density difference to investigate the charge transfer from the adatoms to the $Mo_2P$ substrate. The results shown in Fig. 4 clearly confirm that Li or Na atoms are ionized when they are adsorbed on the surface of single-layer $Mo_2P$. This observation is consistent with the Bader charge analysis results revealing that Li and Na adatoms transfer a charge of 0.984 and 0.983 to the $Mo_2P$, respectively. This highly efficient charge transfer from adatoms to the substrate along with the negative adsorption energy suggesting that the $Mo_2P$ is suitable for the Li and Na storage. It is worthy to note that for both considered adatoms, the second most favourable binding site was found to be on the top of the P atom.

After finding the most stable adsorption site, we increased the coverage of adatoms over the single-layer $Mo_2P$. To this aim, we uniformly increased the metal atoms concentration on the both sides of hexagonal hollow centres until the complete coverage was achieved. In Figs. 4c and 4d, top and side views of the $Mo_2P$ monolayer fully covered by Li and Na adatoms are illustrated. The results clearly confirm that $Mo_2P$ shows extensive stability upon the adatoms converge which is crucial for the anodic application. Notably, Bader charge analysis for the monolayer $Mo_2P$ saturated with Li and Na atoms confirmed close to 99% efficiency in the charge transfer. This means that the charge capacity of the $Mo_2P$ single-layer for both of the Li and Na storage is equal to 240 mAh/g. Directly contributing to the internal electronic resistance and ohmic heat generations during the battery operation, a proper electronic conductivity of an anodic material is highly desirable. To ensure about this issue, we calculated the total electronic density of states for the monolayer $Mo_2P$ with different coverages of Li or Na adatoms. The calculated density of states (Fig.

4e and Fig. 4f) for the all $Mo_2P$ samples with different concentrations of adatoms verify that by increasing the adatoms coverage, the electron states near the Fermi level can be increased. It means that the $Mo_2P$ can yields a high electronic conductivity favourable for the anodic application.

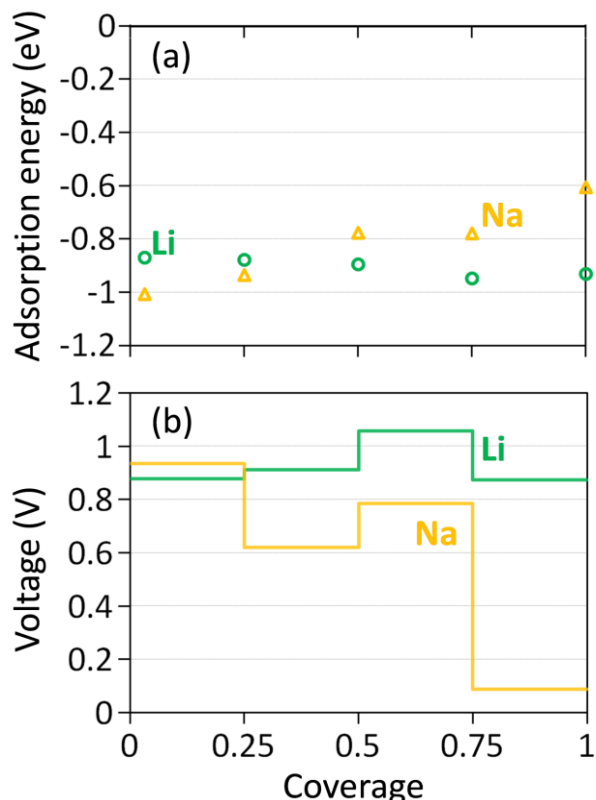

Fig. 5, (a) Evolution of average absorption energy for Li or Na adatoms as a function of coverage ratio. Coverage equal to1 is equivalent with the system that all hexagonal hollow centres are covered on the both sides with the adatoms. (b) Predicted open-circuit voltage profiles as a function of Li or Na adatoms coverage ratio.

We next analysed the evolution of adsorption energy by increasing the adatom converge. Fig. 5a shows the adsorption energy of Li and Na adatoms as a function of the coverage ratio. Interestingly, for Li by increasing the coverage, the adsorption energy slightly increases; while, for Na by increasing the concentration, the adsorption energy may decrease. For both considered metallic adatoms, the adsorption energy remains negative and far from the zero value which is essential for an anode material and further confirms the limited possibility of the adatoms dendrite growth. To choose a material as an anode, the open-circuit voltage profile is another crucial parameter. The open-circuit voltage for the adatom coverages between $x_1 \leq x \leq x_2$ can be estimated by using the following equation [64]:

$$V \approx \frac{(E_{M_{x_1}A} - E_{M_{x_2}A} + (x_2 - x_1)E_M)}{(x_2 - x_1)e} \quad (2)$$

where $E_{M_{x_1}A}$, $E_{M_{x_2}A}$ and $E_M$ are the energies of $M_{x_1}A$, $M_{x_2}A$ and the metallic adatom, respectively. It should be noted that for the accurate estimation of the open-circuit voltage, the Gibbs free energy (G(x) = ΔE + PΔV - TΔS) should be considered. Nevertheless, since PΔV and the entropic term TΔS are negligible, the sole use of the energy term in the voltage estimation is justifiable [36,64]. Fig. 5b shows the voltage profiles of the monolayer $Mo_2P$ as a function of the adatom coverage ratio. As the first observation, the voltage profiles are positive which additionally confirms the suitability of $Mo_2P$ as an anode material for the Li or Na-ion storage. It is worth noting, for the Li adatoms the voltage values are fluctuating around ~1 V, which is primarily due to the slight variations in the adsorption energy. On the other hand, for the Na adatoms, the voltage value starts from 0.94 V at the lowest coverage and reduces to 0.09 V for the fully saturated film. The predicted open-circuit voltage ranges for the $Mo_2P$ are proper for the practical anodic applications.

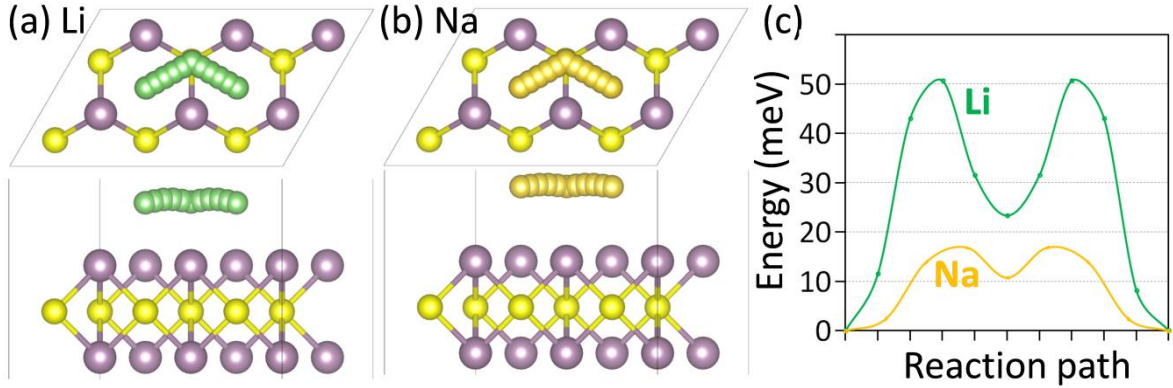

Fig. 6, Top and side views of a single (a) Li or (b) Na adatom diffusion over a single-layer $Mo_2P$. (c) Climbing nudged-elastic band method results for the corresponding diffusion energy barriers.

The diffusion of adatoms is another critical factor characterizing the performance of an anode material, which dominates the charging/discharging rates in rechargeable batteries. A lower diffusion energy barrier results in a higher mobility of ions and improves the charging/discharging rate. In order to explore the diffusion paths and corresponding energy barriers, we employed climbing NEB. Fig. 6 illustrates the top and side views of diffusion pathways of Li and Na atoms over $Mo_2P$. For both Li and Na adatoms, the diffusion follows a zigzag path, through jumping from the hexagonal hollow center to the top of P atoms and then jumping again to the neighbouring hexagonal hollow center. As shown in Fig. 6c, the corresponding energy barriers for the Na and Li adatoms diffusion over the single-layer $Mo_2P$ involves outstanding low diffusion barriers of 51 meV and 17 meV for the Li and Na ions, respectively. The predicted barrier for Li over the $Mo_2P$ is distinctly lower than that for most of 2D materials For example, the Li ions diffusion energy barrier on $Ti_3C_2$ MXene is ~70 meV [65], on phosphorene is 130–760 meV [66], on graphene is ~370 meV [67], on silicene is 230 meV [68], on germanene or stanene is 250 meV [69] and on the flat borophene is 690 meV [31]. Nevertheless, the predicted energy barrier for Li atom diffusion over the $Mo_2P$ is slightly higher than that over $Mo_2C$, 43 meV [36] and $W_2C$, 35 meV [35]. Remarkably, for the Na ions the diffusion energy barrier over $Mo_2P$ is lower than that over both $Mo_2C$ [36] and $W_2C$ [35]. These considerably low diffusion energy barriers over the $Mo_2P$ monolayer confirms its superionic conductivity highly attractive for anodic material applications. As it is clear, $Mo_2P$ satisfies most of the critical characters considered for proper anode materials which owes to its good electronic conductivity, high thermal, dynamical and mechanical stability, ultra low diffusion energy barrier and desirable open-circuit voltage profiles. The main drawback of $Mo_2P$ for the application as an anode lies on its lower charge capacity as compared with commercial anode materials.

## 4. Conclusion

In this work, $Mo_2P$, a novel 2D material was theoretically predicted and investigated. Extensive density functional theory simulations were conducted to study the structural and electronic structures, energetic, dynamical and thermal stabilities, mechanical and optical properties of single-layer and free-standing $Mo_2P$. It was found that $Mo_2P$ can yield a high tensile strength and elastic modulus of 12 GPa.nm and 56 GPa.nm, respectively, and can notably withstand at high temperatures like 1000 K. Our analyses predict the remarkable energetic, dynamical and thermal stability for 2D $Mo_2P$, which confirms the possibility of the synthesis under experimental conditions.

The PBE and HSE06 electronic structure calculations confirm that $Mo_2P$ includes metallic characteristics with the Fermi energy locating at the $d$ bands of transition metal Mo. It is found that the dielectric function is anisotropic along in-plane and out-of-plane directions. Furthermore, by considering the intraband transition contribution, the imaginary and real parts of the dielectric function along in-plane directions have a singularity at zero frequency, indicating a metallic character. In contrast, a positive value of static dielectric constant confirms the semiconducting characters for optical spectra along z-axis. The calculations also demonstrate a reflectivity of around ~100% at extremely low energy less than 1 eV, which provides the potential applications in photonic and optoelectronic systems such as transmitting electromagnetic waves devices.

Finally, the application of $Mo_2P$ as an anode material for Li and Na-ion rechargeable batteries was probed. The open-circuit voltages of 0.88–1.06 V and 0.94–0.09 V were predicted for the Li and Na ions storage. Our DFT results reveal superionic conductivity of $Mo_2P$ which yields remarkably low diffusion energy barriers of 50 meV and 17 meV for Li and Na adatoms, respectively, promising to achieve ultrafast charging/discharging rates. It was discussed that 2D $Mo_2P$ satisfies most of the critical characters as anode material, owing to its desirable electronic conductivity, high thermal, dynamical and mechanical stability, ultra low diffusion energy barrier and proper open-circuit voltage profile. Nonetheless, the charge capacity of $Mo_2P$ for both of the Li and Na storage was predicted to be 240 mAh/g, which is a moderate capacity and stays as the drawback for the application as an anode material. Our comprehensive investigations can be also useful for the future experimental and theoretical studies to explore new 2D materials made from phosphorus and transition metals including different atomic structures such as 2H (with ABA stacking sequence) and 1T (with ABC stacking sequence).

## Acknowledgment

BM and TR greatly acknowledge the financial support by European Research Council for COMBAT project (Grant number 615132).